\documentclass[journal=mamobx,manuscript=article]{achemso}

\usepackage[version=3]{mhchem} 
\usepackage{xcolor}
\usepackage[normalem]{ulem}

\author{Marco A. G. Cunha}
\email{mgalvan1@jhu.edu}
\author{Mark O. Robbins}
\affiliation[Johns Hopkins University]
{Department of Physics and Astronomy, Johns Hopkins University, Baltimore, Maryland 21218, United States}

\title[Flow-induced alignment]
  {Effect of flow-induced molecular alignment on welding and strength of polymer interfaces}

\begin{document}

\begin{tocentry}
	\includegraphics[width=0.75\textwidth]{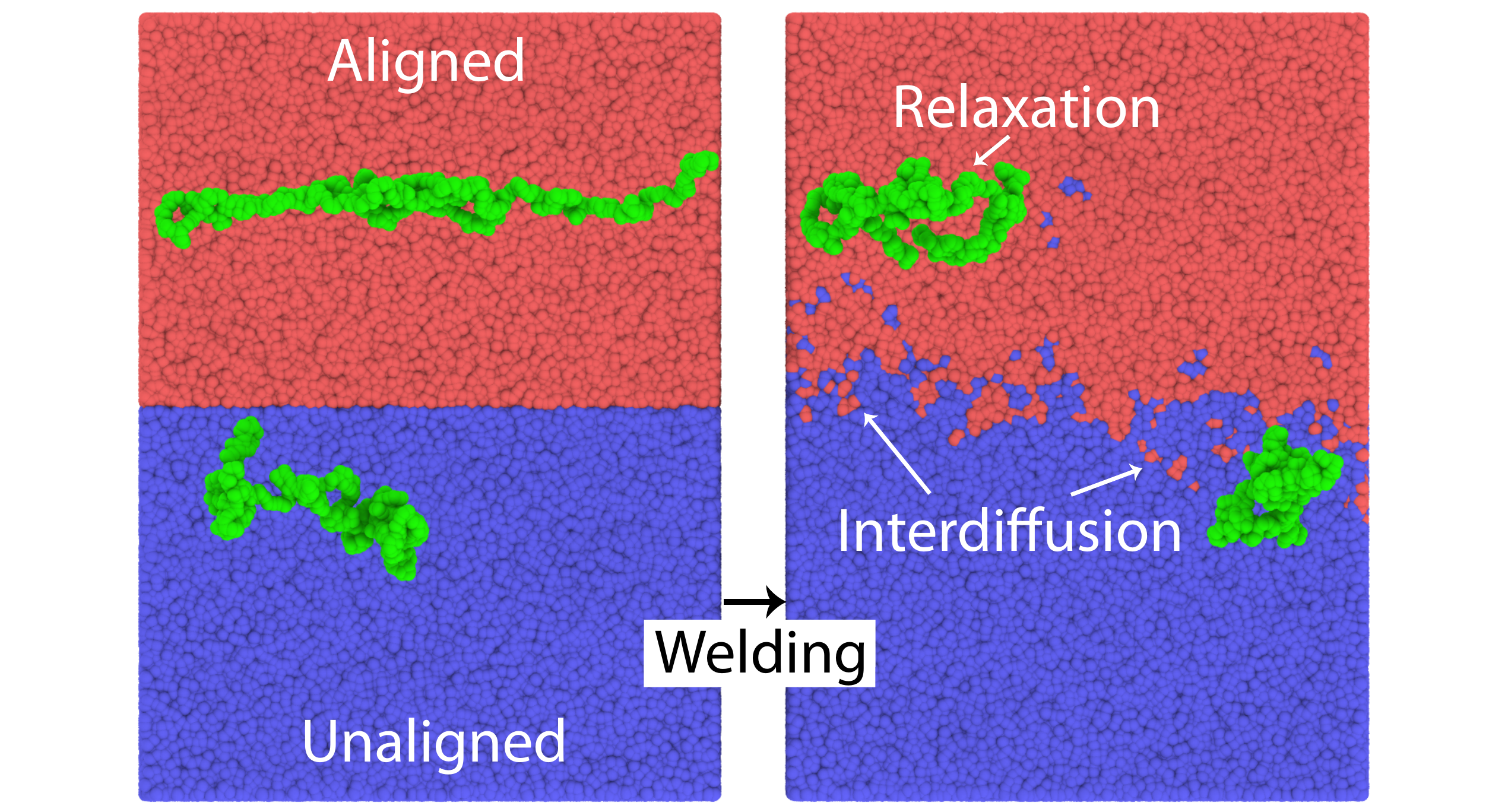}
\end{tocentry}

\begin{abstract}

Structures formed by fused filament fabrication are often substantially weaker than those made with conventional techniques, and fail at the welds between successive layers. One factor that may influence strength is flow-induced alignment of deposited material. Recent work suggests that alignment reduces the entanglement density and thus should accelerate welding by diffusion. Here, coarse-grained molecular simulations are used to test the effect of molecular alignment on diffusion and weld strength. While standard measures show a decrease of the entanglement density with alignment, there is no change in the rate of diffusion normal to the interface or the rate of formation of entanglements across the interface. The time for chain reorientation also remains equal to the equilibrium disentanglement time $\tau_d$. Despite this, simulations of mechanical tests show that welds formed from aligned states are weaker until several $\tau_d$. This is not because the weld itself is weaker, but because aligned material near the weld is weaker than unaligned material. The maximum shear strength and tensile fracture energy of welded systems are the same as bulk systems with the same alignment.
  
\end{abstract}

\section{Introduction}

Welding of polymer interfaces plays a fundamental role in manufacturing processes that range from extruding pipe to additive manufacturing (AM), also known as 3D printing.\cite{wool_polymer_1994,jones_polymers_1999,overeijnder_1983,chua_rapid_2010,n._turner_review_2014} The most common technique used for AM is Fused Filament Fabrication (FFF), where molten polymer is extruded and deposited in successive
layers.\cite{chua_rapid_2010,n._turner_review_2014}
Welding of layers occurs through diffusion of molecules across the interface and stops when the temperature drops below the glass transition temperature $T_g$.
The welding time $t_W$ should be long enough to create strong interfaces,
\cite{seppala_weld_2017,mcilroy_deformation_2017,mcilroy_disentanglement_2017,coogan_prediction_2020}
but short enough that deposited material retains the desired shape.\cite{mackay_importance_2018} 
Tests show that welds remain the point of failure for many printing protocols,\cite{seppala_weld_2017,coogan_bond_2017,n._turner_review_2014,chua_rapid_2010}
leading to great interest in what causes weak interfaces and how they can be strengthened.

One parameter that is affected by deposition protocols is the degree of molecular alignment. Faster deposition and smaller nozzle sizes produce higher shear rates during extrusion.
Bending of the filament as it is deposited produces extensional flow at the bottom of filaments and compressional flow at the top.
When deformation is faster than molecular relaxation times, molecules become aligned along the interface, with greater alignment at the bottom of filaments.\cite{mcilroy_deformation_2017,mcilroy_disentanglement_2017}
Models suggest that alignment reduces the entanglement density
\cite{ianniruberto_convective_2014,baig_flow_2010,milner_microscopic_2001,nafar_sefiddashti_individual_2019}
and thus enhances the dynamics of molecules,
offering the possibility that faster deposition could lead to faster welding.
\cite{mcilroy_deformation_2017,mcilroy_disentanglement_2017}
However, the effect of alignment on welding has not been studied directly.

Recent work shows that coarse-grained molecular dynamics (MD) is a powerful tool for studying weld formation and relating bulk mechanical properties to molecular entanglements, which are not directly observable in experiments.\cite{ge_structure_2013,ge_molecular_2013,ge_tensile_2014,ge_healing_2014}.
Studies of unaligned systems found that
both the tensile fracture energy of the interface $G_I$ and its shear strength
$\sigma_{max}$ scaled with the areal density of entanglements between
chains that started on opposite sides of the interface.
As in experiments, bulk strength was achieved long before the disentanglement or reptation time $\tau_d$ required for molecules to diffuse by their end-end length.
Simulations showed that molecules at the interface only needed to diffuse far enough to form about two entanglements with chains from the other side.\cite{ge_structure_2013,ge_molecular_2013,ge_tensile_2014,ge_healing_2014}.

This paper uses coarse-grained MD to study welding in aligned polymer systems.
Polymer melts are sheared to produce very different degrees of alignment and then brought into contact with equilibrium or aligned substrates.
Although there is an apparent loss of entanglements with increasing alignment,
the rates of relaxation of molecular alignment, of diffusion across the interface, and of formation of interfacial entanglements are relatively insensitive to alignment.
In all cases, the characteristic time scales remain close to equilibrium values
given by tube models of polymer dynamics.\cite{doi_theory_1988}

Despite this, alignment has a pronounced effect on the mechanical properties of the welded interface.
The fracture energy and shear strength rise more slowly than for equilibrium systems, and do not saturate until several $\tau_d$.
Examination of molecular displacements shows that failure does not occur at the weld, but in aligned material nearby.
The strength reduction in welded samples is the same as that for bulk systems with the same alignment.
These results have important implications for the minimum weld time needed to achieve optimum mechanical properties.

The following section provides details of the methods used to align, weld and test welded samples. Then results are presented for the relaxation time of molecular conformations and dynamics at the weld interface.
Next the measured mechanical response to shear and tension is used to determine the shear strength and fracture toughness.
The final section summarizes the results and discusses the implications for printing protocols.

\section{Simulation Model and Methodology}

All of our simulations employed the simulation package LAMMPS\cite{lammps} and used the Kremer-Grest model for linear homopolymers\cite{kremer_dynamics_1990,everaers_kremergrest_2020}.
Van der Waals interactions between monomers of mass $m$ are modeled by a Lennard-Jones (LJ) potential with interaction energy $u_0$ and monomer diameter $a$.
To accelerate simulations the potential is truncated and shifted to zero at a cutoff radius $r_c$. Mappings to real polymers \cite{everaers_kremergrest_2020} typically give $a \sim 0.5$ nm and the unit of stress $u_0/a^3 \sim 50$ MPa.

Linear chains of length $N = 500$ were made by connecting beads to their nearest-neighbors with a bond potential.
All melt simulations including equilibration, shearing and welding were performed with the usual finitely extensible nonlinear elastic (FENE) potential \cite{kremer_dynamics_1990}.
This prevents chains from crossing or breaking, but 
chain scission plays an important role in mechanical failure of glassy polymers.\cite{rottler_cracks_2002,ge_structure_2013,ge_molecular_2013, ge_healing_2014,ge_tensile_2014}
As in past studies of welding,\cite{ge_structure_2013,ge_molecular_2013,ge_healing_2014,ge_tensile_2014} we used a breakable quartic potential in simulations of solid deformation.
The potential is chosen to have nearly the same bond length as the FENE
potential and to break at a force corresponding to about 100 times the force necessary to break LJ bonds,
as this ratio has been found to be representative of real polymers.\cite{rottler_cracks_2002,ge_healing_2014}
The entanglement length for this model is $N_e = 85 \pm 7$ and polymers with $N = 500$ are long enough to have the mechanical response of highly entangled polymers.\cite{ge_healing_2014}

Melt slabs with $M = 1200$ chains were equilibrated at temperature $T = 1.0u_0/k_B$ and density $\rho=0.85 a^{-3}$ with $r_c = 2^{1/6}a$ following the standard double-bridging algorithm.\cite{auhl_equilibration_2003} The samples were periodic in the $x-$ and $y-$directions with periods $L_x=185.2 a$ and $L_y= 61.7 a$, while the length in the confined $z-$direction, $L_z \approx 61.7 a$, was fixed by using two repulsive featureless walls.\cite{ge_molecular_2013}
Small deviations from equilibrium behavior are present close to the walls, which is consistent with reports in the literature.\cite{kirk_chain_2017,kirk_entanglement_2019}
We also equilibrated melts with $M=2400$ chains using the same boundary conditions but double the length in the $z-$direction.
These systems had the same geometry and were exposed to the same shear history as welded samples.
Since they had no initial interface, we will refer to them as bulk samples and use mechanical tests on
them as a reference for the intrinsic response of bulk aligned systems.

As noted in the introduction, FFF deforms filaments in complicated ways.
There is shear flow in the nozzle and the nozzle may scrape over the deposited filament, causing further shear. There is also elongational (compressional) flow at the bottom (top) surface as the filament bends and joins the growing part, and some relaxation of chains will occur during this process.
Rather than attempting to model this process in detail, we consider systems where the system has been sheared at rates that produce dramatic changes in the degree of alignment and examine the trends with alignment. 

Slabs were sheared along the $x-$direction with a velocity gradient along the $z-$direction.
Slabs are finite in the $z-$direction and shear was imposed by applying constraints on atoms within a distance $5a$ of the confining repulsive walls.
Applying a force in the $x-$direction or constraining the average velocity $\langle v_x \rangle$ of atoms in the $x-$direction produced statistically equivalent chain conformations and weld strength and we show results for the velocity constraint below.
The constraints produced a linear velocity profile in the center of the film with shear rate $\dot{\gamma} \equiv \partial v_x/\partial z$.
As discussed in the next section, the alignment, entanglement density, and relaxation dynamics in this region were consistent with bulk simulations of shear at the same $\dot{\gamma}$ with periodic boundary conditions in all three directions.

During shear, the temperature was kept at $T = 1.0u_0/k_B$ with a Langevin thermostat.
To prevent any bias of the flow profile, only the deviation from a linear
velocity profile was thermostatted.
The damping time was $1 t_0$,
where $t_0 \equiv \sqrt{m u_0}/a$ is the characteristic Lennard-Jones time.
Decreasing the damping by an order of magnitude did not produce statistically significant changes in velocities, stresses or chain conformations.
The entanglement time for this potential,
$\tau_e= 1.07 \pm 0.21 \times 10^4 \ t_0$, has been determined in past studies.\cite{svaneborg_characteristic_2020}
Then scaling arguments\cite{Likhtman2002} predict a disentanglement time $\tau_d = 1.2 \times 10^6 \ t_0$ that is consistent with the measured relaxation time for alignment, as discussed below.
The time for chains to relax their length in the tube is the Rouse time $\tau_R = Z^2 \tau_e = 3.7 \times 10^5 \ t_0$, with $Z=N/N_e = 5.9$.
Steady-states were reached after shearing for a time between $10^6 \ t_0$ and $5 \times 10^6 \ t_0$, depending on the rate. 
This is about $\tau_d$ or longer and corresponded to strains of at least 100. The system was determined to be in steady state by looking at both the shear stress and chain statistics.
In the following we will use the shorthand $1 \ Mt_0 \equiv 10^6 t_0$.

The degree of chain stretching during shear
is determined by the competition between shear deformation and chain relaxation in the tube.\cite{larson_structure_1998,doi_theory_1988}
This is characterized by the dimensionless Rouse-Weissenberg number:
$Wi_R = \dot{\gamma}\tau_R$.
	Results are presented for $\dot{\gamma}=2 \times 10^{-5} \ t_0^{-1}$, $ 2 \times 10^{-4} \ t_0^{-1} $ and $ 1 \times 10^{-3} \ t_0^{-1}$, corresponding to
	$Wi_R \approx 7.4$, 74 and 370.
The lowest $Wi_R$ is comparable to the highest Weissenberg numbers estimated in recent models of FFF deposition, and produces chain stretches of the same order.\cite{mcilroy_deformation_2017,mcilroy_disentanglement_2017}.
However, this $Wi_R$ produces relatively little stretching and loss of entanglements in our simulations and past work.\cite{baig_flow_2010,nafar_sefiddashti_individual_2019}
Higher shear rates were studied to explore the range of possible
alignments and entanglement loss.

\begin{figure}
  \includegraphics[width=\columnwidth]{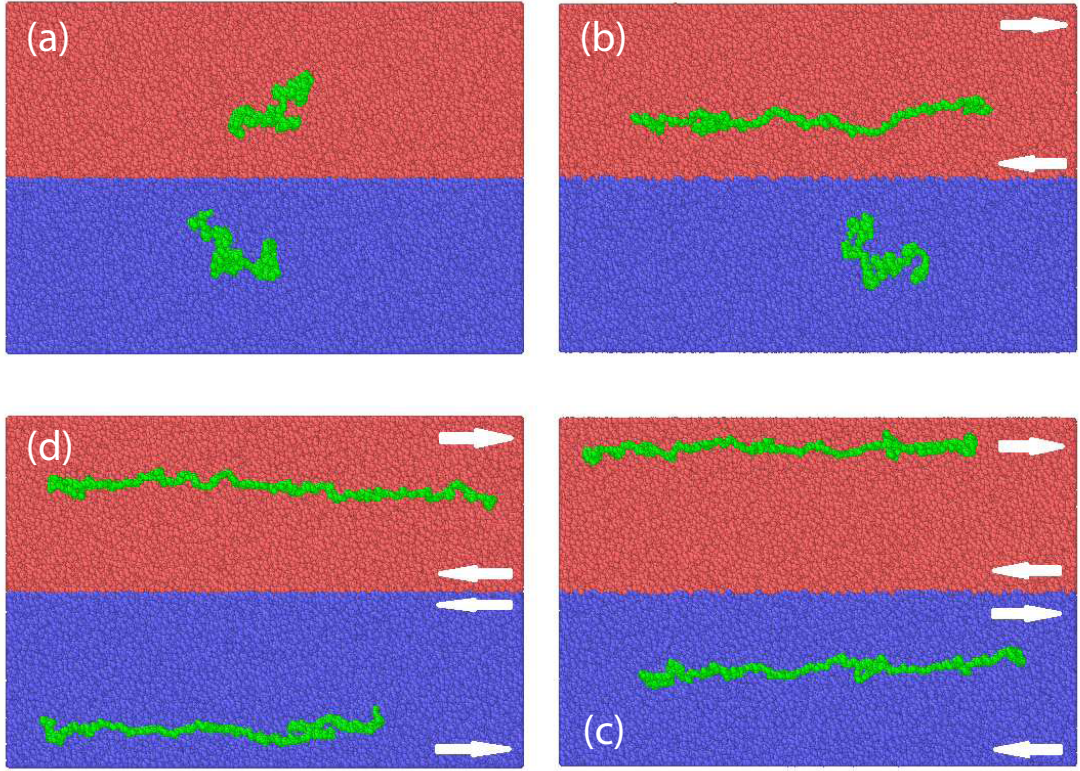}.
	\caption{Different configurations used for welding simulations:
	(a) Eq/Eq - two equilibrated slabs, (b) Eq/Sh - equilibrated slab below sheared slab, (c) Sh/Sh - two slabs sheared in the same direction, and (d) Sh/Sh$_{rev}$ - slabs sheared in opposite directions.
	Arrows indicate the direction of the shear velocity and $Wi_R=74$.
	Representative chains are shown in green to indicate the degree of alignment.
	}
  \label{fgr:systems}
\end{figure}

After equilibration or shearing to steady state, two slabs were moved into contact and allowed to weld.
The repulsive walls between slabs were removed and they were moved as close as possible without having any repulsive interactions between monomers from different slabs.
This produces an interface that is parallel to the $xy-$plane and
welding occurs by interdiffusion along the $z-$direction.
As depicted in Fig. \ref{fgr:systems},
four types of system were prepared for welding. The reference configuration
has both slabs in equilibrium. The second configuration has only one surface sheared, since the top of a filament has more time to relax and is under slower deformation than the bottom.\cite{mcilroy_deformation_2017,mcilroy_disentanglement_2017} The final two configurations have slabs sheared in the same or opposite directions to mimic different possible printing patterns.

Welding was allowed to occur for the desired weld time $t_W$ before quenching into the glassy phase.
Bulk systems were allowed to relax for the same time intervals from the same $Wi_R$ to determine how interfaces changed the mechanical response.
The temperature during welding will not be constant, but an effective
$t_W$ at a reference temperature can be obtained from the thermal history and time-temperature superposition.\cite{larson_structure_1998,mcilroy_deformation_2017,mcilroy_disentanglement_2017,seppala_weld_2017,coogan_prediction_2020}

After welding or bulk relaxation, systems were quenched from the melt temperature of $T = 1.0\ u_0/k_B$ to $T = 0.2\ u_0/k_B$, which is below the glass transition temperature $T_g \approx 0.35\ u_0/k_B$ for this system. The cooling protocol followed past work.\cite{ge_structure_2013,ge_tensile_2014} First the Lennard-Jones cutoff was increased to $r_c = 1.5\ a$. Then the system was quenched at constant volume at a rate of $\dot{T} = -10^{-3}\ u_0/k_B t_0$ to about $T=0.5\ u_0/k_B$ where $P$ has dropped to nearly zero.
The rest of the quench was done at $P = 0$ at a rate of $\dot{T} = -2 \times 10^{-4}\ u_0/k_B t_0$. $P_{xx}$ and $P_{yy}$ were controlled with a Nose-Hoover barostat with damping time $50\ t_0$ and $T$ was maintained with a Langevin thermostat with time constant $1\ t_0$.
After the quench, the bonding potential along the chain was changed from FENE to the breakable quartic potential used in previous weld simulations.\cite{ge_structure_2013,ge_molecular_2013,ge_healing_2014,ge_tensile_2014}.

To deform the resulting glasses, we imposed rigid displacements on
beads within $5a$ of the walls at the top and bottom of the system that are farthest from the welded interface.
Stress was determined by monitoring the total force imposed on these groups of atoms.
The temperature during deformation was kept at $T = 0.2\ u_0/k_B$ by a Langevin thermostat with a damping time of $1 \ t_0$ and 
acting only in directions orthogonal to deformation in order to avoid biasing displacements.

For simulations of tensile fracture of the welded interface, we followed past studies of the fracture energy of glassy polymers.
\cite{rottler_cracks_2002,ge_tensile_2014}. 
The top and bottom were moved apart at a constant velocity $v = 0.01 \ a/t_0$ in a direction perpendicular to the interface.
Simulations of shear fracture mimiced shear tests of lap joints.\cite{wool_polymer_1994}
A shear displacement was applied in the weld plane and normal to the printing direction along which molecules are aligned. This orientation was chosen to most closely reflect the shear in tear tests on printed samples.\cite{seppala_weld_2017}
Top and bottom were displaced in opposite directions along the y axis, with a velocity consistent with a strain rate of $2 \times 10^{-4} \ t_0^{-1}$. This is the same strain rate used in past simulations \cite{ge_structure_2013,ge_molecular_2013,rottler_shear_2003}, where it was determined to be small enough to allow the stress to equilibrate across the system.

Entanglements are known to have a strong effect on the mechanical properties of
bulk glassy polymers\cite{ward_2012,haward_1997} and welds.\cite{wool_polymer_1994}
We use both Primitive Path Analysis (PPA)\cite{everaers_rheology_2004,sukumaran_identifying_2005,zhou_primitive_2005} and the Z1 code \cite{kroger_shortest_2005,hoy_topological_2009} to identify entanglements between chains.
Results are obtained from the entire volume of a slab,
but excluding the region near the surfaces had little effect.

An executable for the Z1 code was provided by Martin Kr\"oger. Several versions of the PPA have been proposed. We used one where the attractive FENE potential has a constant attractive force below some separation to produce a constant tension along the chain.\cite{zhou_primitive_2005}
The influence of excluded volume is reduced by shrinking the radius of monomers by a factor of 8 and adding 7 extra beads between each initial monomer to prevent chain crossing.
In both approaches, topological constraints (TCs) are identified as contacts (repulsive interactions) between chains after they have been shortened as far as possible with ends fixed and no chain crossing.
The Z1 code determines $Z$ from the number of points along a chain where it changes direction by more than a threshold angle due to contact with another chain or chains.
For PPA we count the number of connected regions where a chain contacts other chains.
Gaps in contact of less than $a$ are excluded to avoid double counting contacts with multiple chains at the same point.
Values of $Z$ produced in this way are similar for the two methods.
However, they are larger than the number of entanglements determined from rheological measurements because there are multiple TCs within the Kuhn length of the primitive path.\cite{hoy_topological_2009,kroger_shortest_2005,zhou_primitive_2005,everaers_rheology_2004}
For $N_e=85 \pm 7$ and $N=500$ there should be $N/N_e -1 = 4.9\pm0.5$ entanglements per chain, but we find about twice this many TCs in equilibrium $Z \approx 12 $ from the Z1 code and $Z \approx 11 $ from PPA.

\section{Results}

\subsection{Structure and Entanglements During Steady Shear}

\begin{figure}
  \includegraphics[width=0.5\linewidth]{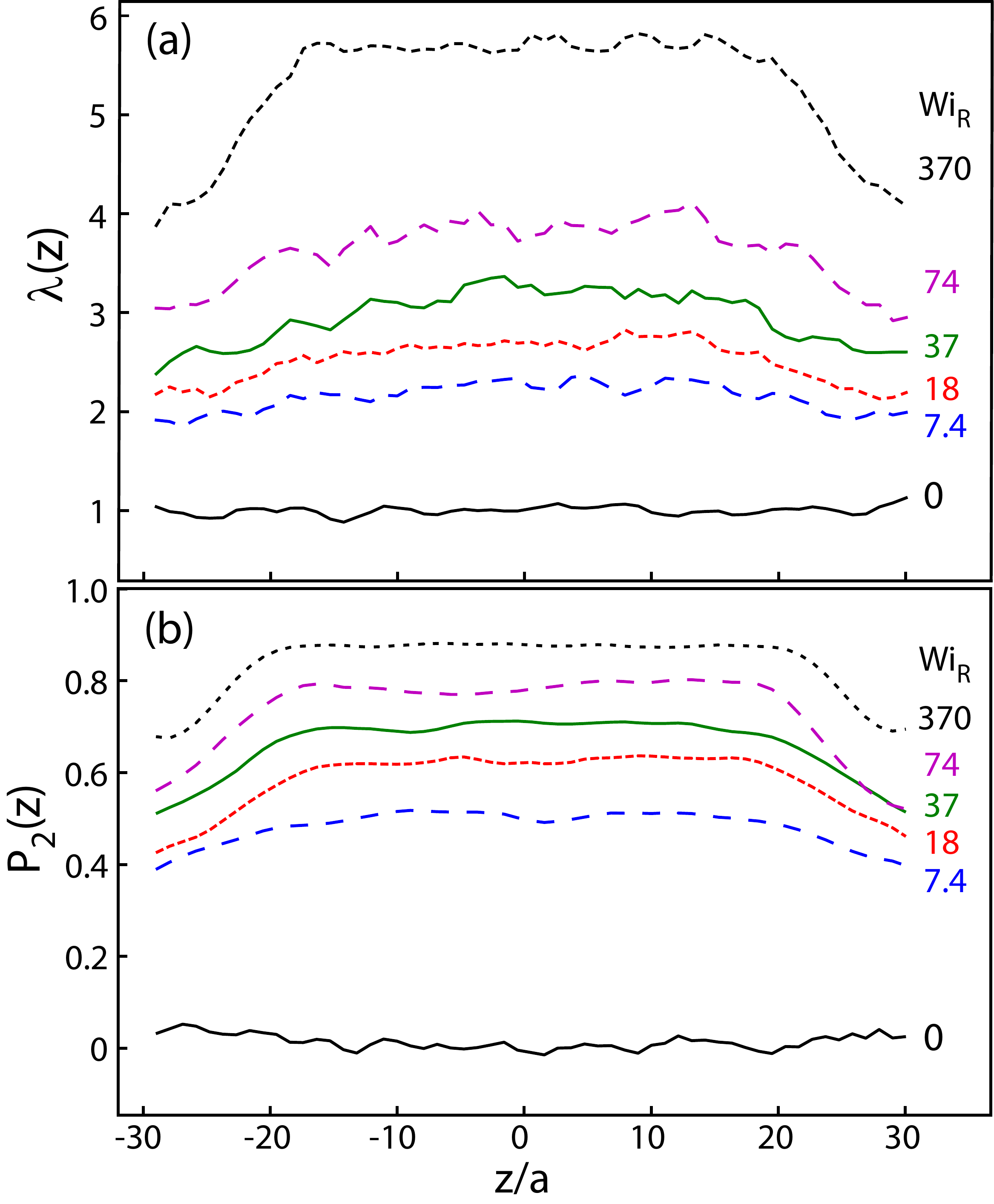}
	\caption{(a) Height dependence of the stretch $\lambda(z)$ of the end-end length of chains in steady-state shear at the indicated $Wi_R$. Data are binned by the height of chain ends and repulsive walls are located at $z/a = \pm 30$.
	(b) Nematic order along flow, $P_2(z)$, of segments of length $N_e$ at the same $Wi_R$.
	Data are binned by segment ends.
	Sytems were sheared at $\dot{\gamma} t_0=2 \times 10^{-5}$, $5 \times 10^{-5}$, $1 \times 10^{-4}$, $2 \times 10^{-4} $ and $ 1 \times 10^{-3} $, corresponding to $Wi_R \approx 7.4$, 18, 37, 74 and 370.
	}
  \label{fgr:ree_z}
\end{figure}

To quantify the degree of alignment produced by steady-state shear of melts, the stretching and orientation of chains were calculated as a function of height $z$ along the velocity gradient direction.
Figure \ref{fgr:ree_z} shows the degree of stretch $\lambda (z)$, defined as the ratio of the root-mean-squared (rms) end-end length of chains $\langle R_{ee}^2 \rangle^{1/2}$ to the equilibrium value of $28.5 \ a$.
Results were binned by the location of each end to obtain local information about chains starting near the surfaces.
The equilibrium results show $\lambda$ remains near unity across the slab.
For $Wi_R=7.4$, $\lambda \approx 2$, with small decreases in stretching near the surface.
Surface effects become more apparent as $Wi_R$ increases, and extend up to about $10 \ a$.
Nonetheless, the stretch at each $z$ increases substantially with increasing rate,
and these sheared systems provide initial states with a broad range of stretches at the interface where welding will take place.

Theories of polymer rheology often quantify stretch by taking the trace of the conformation tensor
$ A_{ij} \equiv \langle {R_i R_j}\rangle /2R_g^2$, where $R_i$ are components of the end-end vector, and the equilibrium value of $\langle R_i^2 \rangle $ is twice the square of the equilibrium radius of gyration $R_g$.
Deviations from equilibrium are quantified by $tr {\bf A}-3 = 3(\lambda^2-1)$.
In the following we focus on $Wi_R=7.4$, $74$ and $370$, corresponding to surface values of $\lambda \approx 2$, 3 and 4 and $tr{\bf A}-3 \approx 9$, $24$ and $45$, respectively.
McIlroy and Olmsted found values of $tr{\bf A}-3 \sim 8$ were produced for one set of printing parameters, which is comparable to our lowest rate. Our results for the largest $Wi_R$ correspond to rates about 50 times higher.

Entanglements play a critical role in the mechanical properties of glassy polymers.\cite{ward_2012,haward_1997,wool_polymer_1994}
To examine the degree of chain alignment on the entanglement scale, we divided chains into segments of length $N_e$ with arbitrary starting points relative to the chain ends.
The degree of orientation along the flow direction was then quantified
by calculating the nematic order parameter:
\begin{equation}
    P_2 \equiv \langle3\cos^2\theta-1\rangle/2 \ \ ,
\end{equation}
where $\theta$ is the angle between the $x-$axis and the vector between segment ends.
The value of $P_2$ is zero for random orientation and reaches one for perfect alignment. 
Once again, results for $Wi_R=0$ and $7.4$ are relatively constant across the slab.  While there are drops in $P_2$ near the wall for larger $Wi_R$,
the degree of order at the surface rises significantly from about 0.4 to 0.7.
Similar plots for the orientation of the entire chain show $P_2$ rising from 0.6 to 0.8 for the same $Wi_R$.

\begin{figure}
  \includegraphics[width=0.6\linewidth]{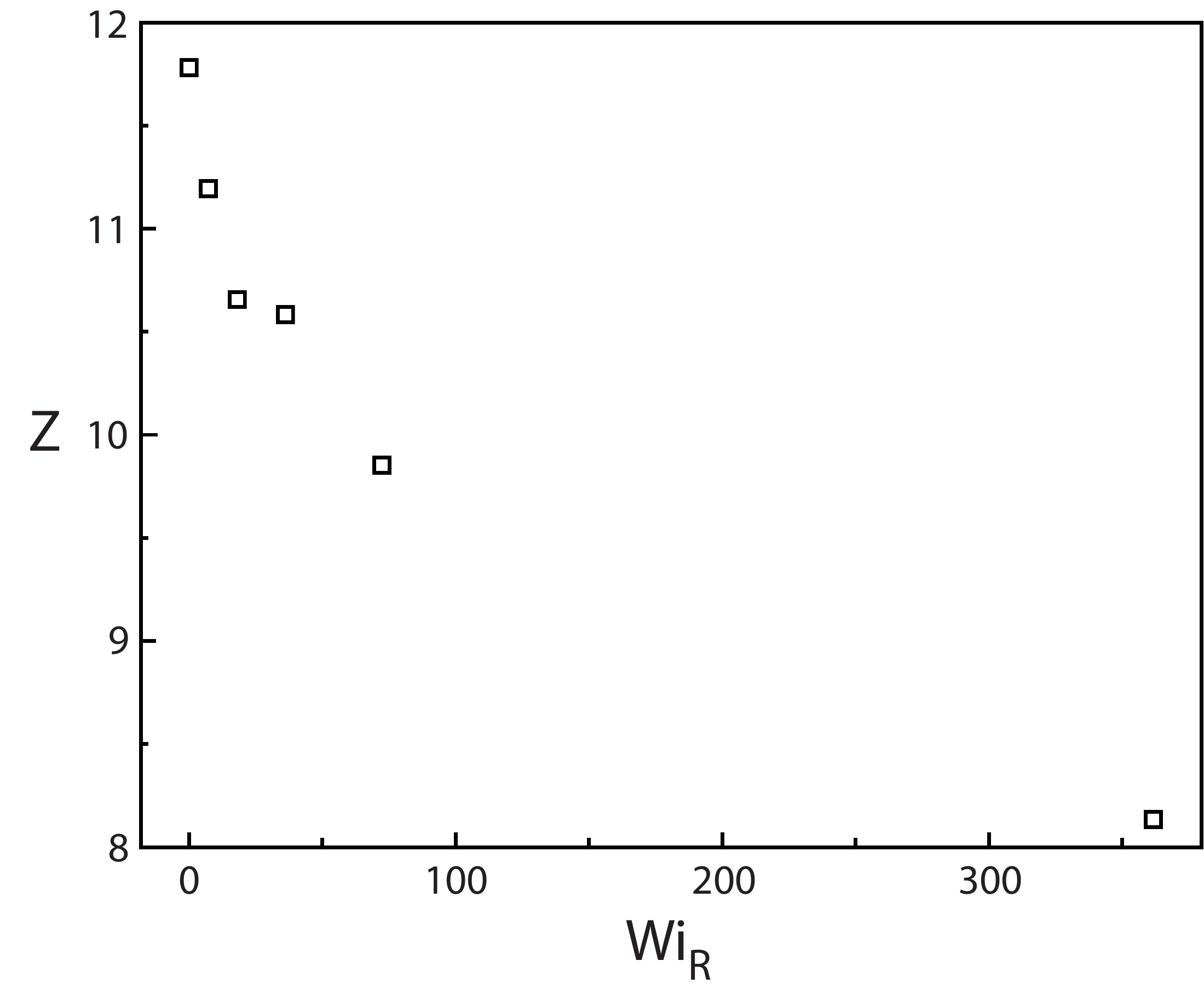}.
	\caption{Number of entanglements per chain, $Z$, from Z1 code as a function of Rouse Weissenberg number $Wi_R$. Statistical errors are of order 0.2.
	}
  \label{fgr:Zgd}
\end{figure}

A variety of evidence suggests that alignment reduces the number of entanglements.\cite{ianniruberto_convective_2014,baig_flow_2010,nafar_sefiddashti_individual_2019,oconnor_stress_2019}
Indeed models of FFF predicted more than 60\% of entanglements were lost at
alignments corresponding to our lowest rates.\cite{mcilroy_deformation_2017,mcilroy_disentanglement_2017}
Figure \ref{fgr:Zgd} shows the rate dependence of $Z$ in steady-state flow from the Z1 code.
There is a drop by about 20\% at $Wi_R=74$ and 30\% at $Wi_R=370$.
Slightly larger drops
have been seen in simulations of polyethylene at these $Wi_R$.
\cite{baig_flow_2010,nafar_sefiddashti_individual_2019}
There, and in our ongoing studies of semiflexible chains, the drop in $Z$
at a given $Wi_R$ rises with increasing $Z_{eq}$. 
Elongational flow, which may be present as the filament bends during extrusion, can lead to larger reductions in $Z$
from the Z1 code.\cite{oconnor_stress_2019}

In contrast to Fig. \ref{fgr:Zgd}, $Z$ drops by less than 5\% when measured with PPA for the most aligned states.
This stark difference in behavior persists for all the treatments of multiple contacts that we tried.
One possibility is that because chains are much straighter, the angle-change criterion used in the Z1 code excludes more contacts as $Wi_R$ increases.
The origin of this difference will be the subject of future studies, but is not central to the development of weld strength considered here.
The results below and in recent work \cite{oconnor_stress_2019} show that whether reductions in $Z$ are apparent or real,
the time scales that control relaxation retain their equilibrium values in highly aligned states and alignment does not change the rate of interdiffusion at welds.

\subsection{Relaxation and Diffusion During Welding }

After shearing is stopped, the stretch and orientation of chains begin to relax from
the steady-state values described in the previous section.
In our simulations, shear stops at the same time that slabs are placed in contact and begin to weld through interdiffusion.
Thus the time for relaxation equals the weld time $t_W$ and all results are presented
in terms of $t_W$.
In FFF there might be several stages of deformation and relaxation before
a filament contacted the substrate.
The main effect of such stages is just to alter the alignment at the time
of contact and the onset of welding.
This section shows that the rates of relaxation and interdiffusion are
quite insensitive to alignment.

\begin{figure}
  \includegraphics[width=0.6\columnwidth]{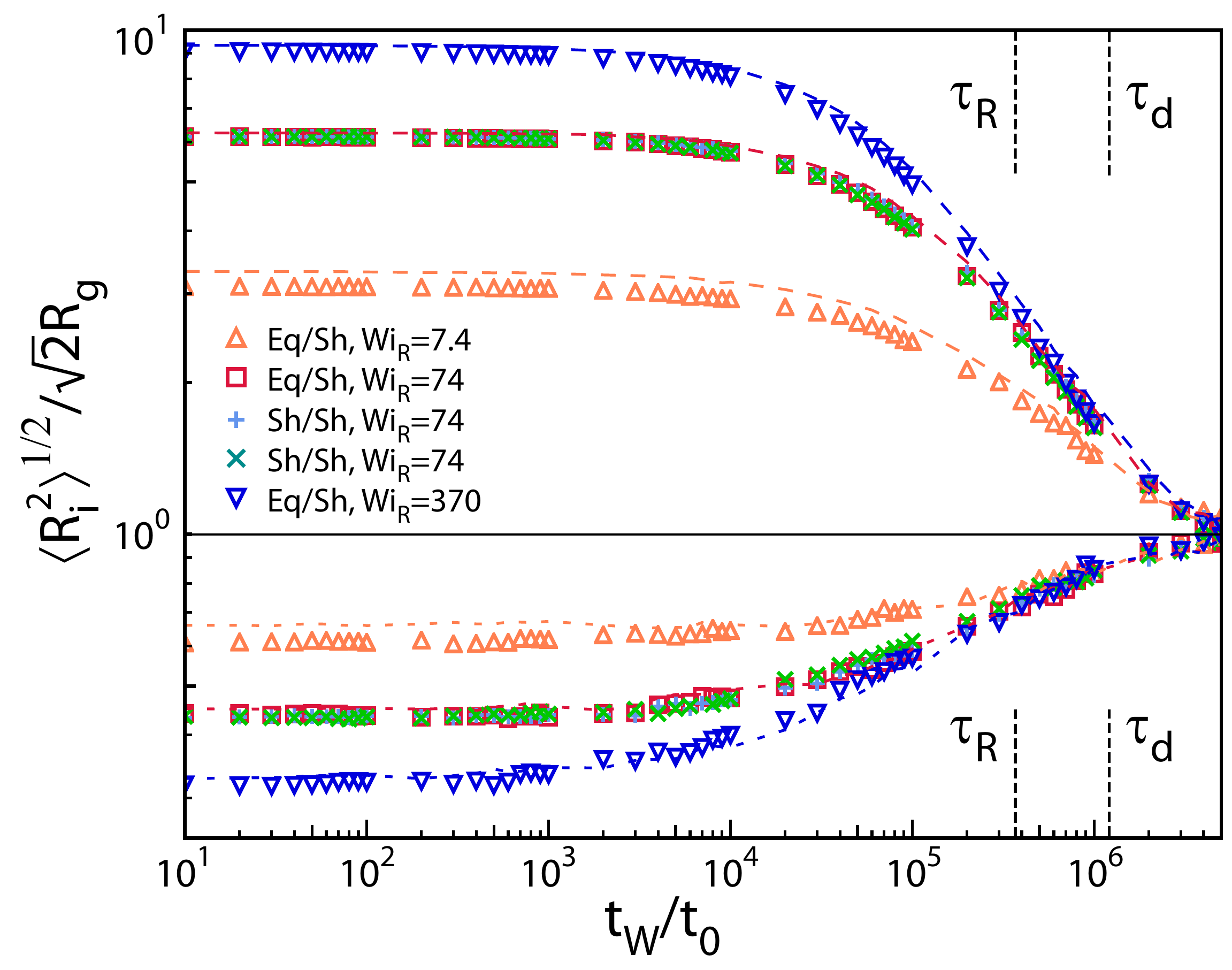}.
	\caption{
		Relaxation of rms components $R_i$ of chain end-end vectors normalized by the equilibrium value $\sqrt{2} R_g$ for the indicated configurations and initial shear rates.
Results for the $x$ ($z$) component are above (below) unity.
Symbols show results for welded systems and broken lines with the same color are for bulk systems at the same time after shear stopped.
Vertical dashed lines indicate $\tau_R$ and $\tau_d$.
Only chains in slabs that were sheared were included in the average.
	}
  \label{fgr:ree}
\end{figure}

Figure \ref{fgr:ree} shows the relaxation of the rms components of the end-end vector along the flow direction (x) and normal to the interface (z).
Results for the $y-$ direction are similar to those along $z$.
Results are averaged only over chains that started in sheared slabs and are normalized by $\sqrt{2} R_g$ so that they reach unity in equilibrium.
Results for the bulk systems without an interface are shown by lines for comparison.
The magnitude of changes in Fig. \ref{fgr:ree} increases with $Wi_R$, but the time scale for change is always of order $\tau_R$.

Immediately after shear is stopped, almost all of the end-end length is associated with the projection along the $x-$axis.
Thus this component is stretched by about $\sqrt{3} \lambda$.
Little change along $x$ is seen until times of order $0.1 \tau_R $,
and by $\tau_R$ all curves have dropped significantly.
The other components start to increase at smaller times as $Wi_R$ increases,
but have also undergone most of their relaxation by $\tau_R$.
These results are qualitatively consistent with models for the linear response of polymers.\cite{larson_structure_1998,doi_theory_1988}
They predict that stretch relaxes with a distribution of relaxation times that correspond to different segment lengths along the chain.
The longest relaxation time, $\tau_R$, is associated with the end-end scale.

\begin{figure}*
  \includegraphics[width=0.6\columnwidth]{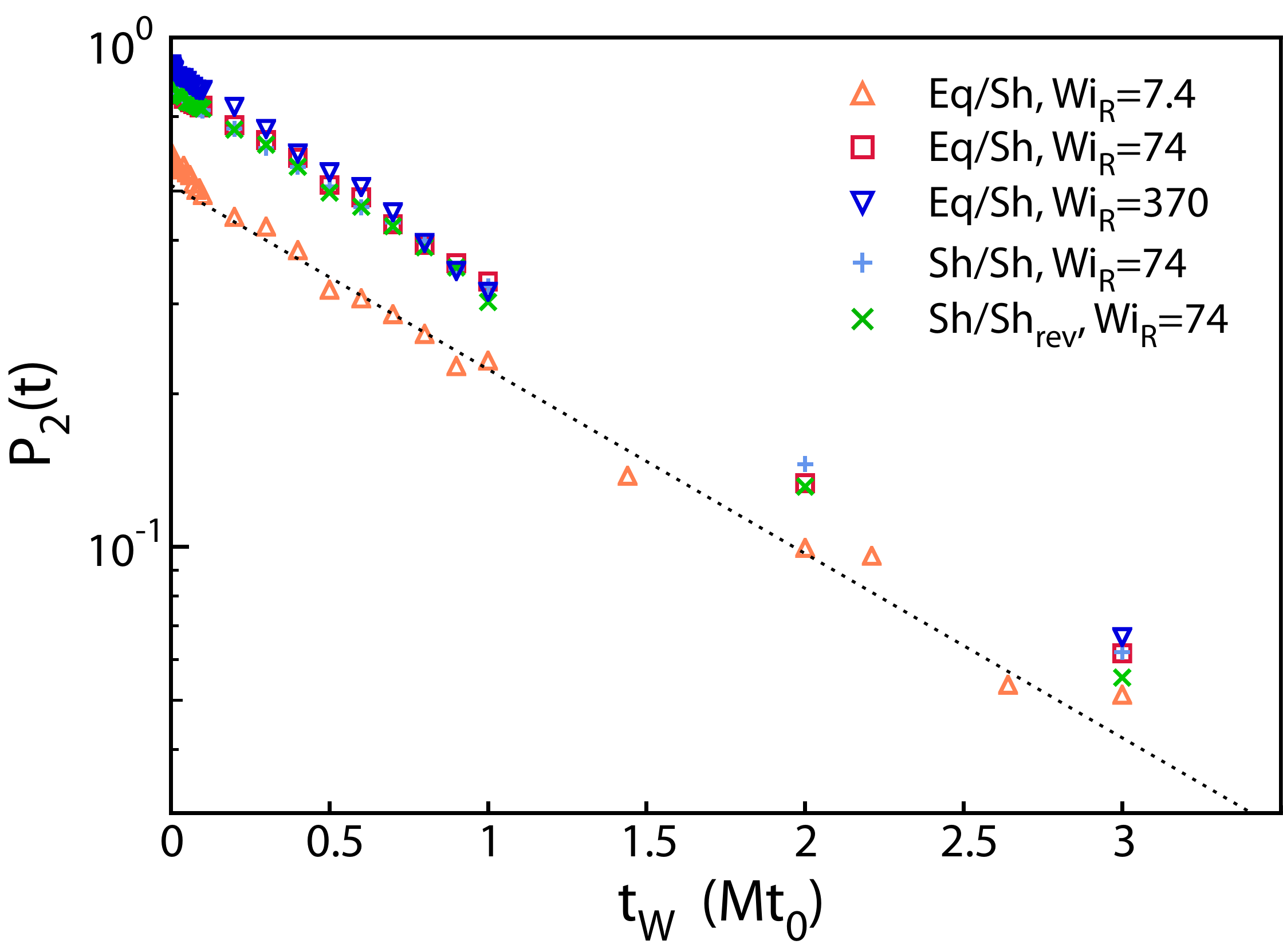}.
  \caption{Relaxation of the nematic order parameter on the end-end scale with time in $M t_0 \equiv 10^6 t_0$, for different configurations and shear rates.
	The initial alignment depends on $Wi_R$, but all systems relax exponentially at a rate consistent with the equilibrium $\tau_d = 1.2Mt_0$ (dotted line).
	Only chains in slabs that were sheared were included in the average.
	}
  \label{fgr:P2t}
\end{figure}

The same models for linear response predict a simple exponential relaxation of the nematic order parameter on the end-end scale:
$P_2 (t) \propto \exp{(-t/\tau_d)}$.
Figure \ref{fgr:P2t} shows that our data follow this behavior even though chains are very far from equilibrium.
Results for different $Wi_R$ start with different degrees of order, but
plots of $\log(P_2)$ against time have a common slope that is consistent with the equilibrium $\tau_d$.


\begin{figure*}
  \includegraphics[width=0.45\linewidth]{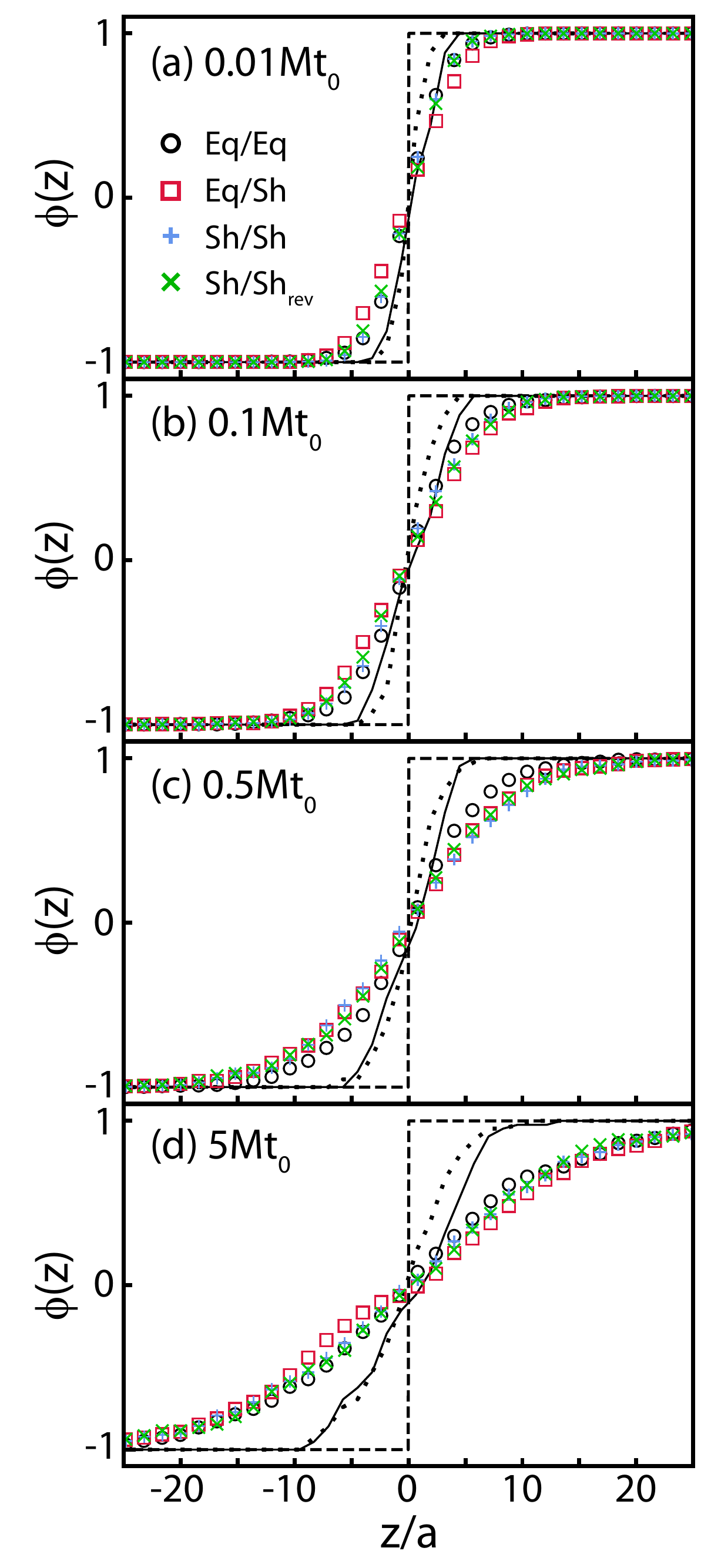}.
	\caption{Concentration profiles at welding times (a) $t_W= 0.01 Mt_0 \sim \tau_e$, (b) $t_W=0.1 \ Mt_0 = 0.27 \tau_R$, (c) $t_W=1 \ Mt_0=2.7\tau_R$ and (d) $t_W=5 \ Mt_0 =13.5 \ \tau_R \sim 4 \ \tau_d$. The profile is a step function at $t_W=0$ (dashed line) and broadens through diffusion and interface roughening. Black dotted (Eq/Eq) and solid (Eq/Sh) lines show the contribution $\phi_0$ from interface roughness as described in the text.  Sheared systems had $Wi_R=74$ and concentrations were binned over $1.5 \ a$ in the z-direction.
}
  \label{fgr:conc}
\end{figure*}

One way of quantifying interdiffusion is through the evolution of the concentration profile.
Molecules starting on the top of the system are labelled type 1 and molecules on the bottom are type 2. They have number density $\rho_1(z)$ and $\rho_2(z)$,
respectively.
The concentration profile
\begin{equation}
	\phi(z) \equiv (\rho_1(z)-\rho_2(z))/(\rho_1(z)+\rho_2(z))
\end{equation}
starts as a step function from -1 to +1 at $z=0$ and broadens as molecules diffuse.
Figure \ref{fgr:conc} shows $\phi(z)$ for $Wi_R=74$ at times ranging from $\tau_e$ to $4 \ \tau_d$, when memory of alignment has been lost.
In each case, the top particles were in a sheared slab.
These times are chosen because they represent important stages in the evolution of the mechanical strength of welds (see below).

All concentration profiles in Fig. \ref{fgr:conc} are approximately antisymmetric, $\phi(-z) \approx -\phi(z)$.
This symmetry would be broken in the Eq/Sh system if diffusion were
faster in aligned systems.
Faster diffusion would also lead to a more rapid broadening of the concentration profiles in sheared systems,
but differences between the four types of starting state remain small over the entire range of times.
The largest differences are observed at intermediate times where the Eq/Eq profile is narrower, suggesting slower interdiffusion.
While this would be consistent with entanglement loss accelerating interdiffusion for aligned systems, 
closer examination shows that the difference comes from changes in interface roughness.
As highly stretched chains relax their length along $x$, they expand along $z$.
Local fluctuations deform the interface and the resulting roughness broadens the concentration profile without mixing molecules that started on different sides of the interface.

To determine variations in the local interface height, concentration profiles $\phi(z)$ were calculated within square bins of width $10 \ a$ in the $x-y$ plane.
The height $z_0(x,y)$ where $\phi =0$ was determined by fitting $\phi$ to $\tanh (z-z_0(x,y))$.
If the local concentration profile remained a step function, but $z_0$ varied,
the concentration profile would broaden to $\phi_0(z) = 2\int_{-\infty}^z dz_0 p(z_0) -1$ where $p(z_0)$ is the probability of finding a local interface height $z_0$.
Black dotted and solid lines in Fig. \ref{fgr:conc} show $\phi_0$ for the Eq/Eq and Eq/Sh ($Wi_R=74$) systems, respectively.
At early times, $\phi_0$ is only slightly narrower than $\phi$,
indicating that surface roughness plays an important role.
The larger width of $\phi_0$ for Eq/Sh than Eq/Eq systems, shows the corresponding broadening of $\phi(z)$ comes from 
roughness rather than faster diffusion.
By $t_W =5 Mt_0$, $\phi_0$ is much narrower than $\phi$ for all systems, implying that diffusion dominates over roughness

Past experimental and theoretical work on unaligned systems showed a correlation between interfacial strength and the distance monomers from one side have moved into the other melt.\cite{wool_polymer_1994,ge_structure_2013,ge_molecular_2013,ge_tensile_2014}.
The interdiffusion distance is defined as 
$\langle d_1\rangle = \int_{0}^{\infty} z \rho_1(z)dz/\int_{0}^{\infty} \rho_1(z)dz$ for type 1 monomers.
The definition is analogous for type 2 monomers, with the integration limits changed to $-\infty < z < 0$.
The roughness identified above will contribute to $\langle d_i \rangle$, but does not reflect mixing of species that leads to mechanical strength.
To eliminate roughness, we calculated $\langle d \rangle \equiv 0.5 (\langle d_1 \rangle + \langle d_2 \rangle)$ using $z_0(x,y)$ as the limit of integration instead of zero.
To facilitate comparison with past work,\cite{ge_structure_2013,ge_molecular_2013,ge_tensile_2014} $z_0$ was calculated for bins of width $10 \ a$, which
is comparable to the equilibrium $R_g \approx 11.6 \ a$. 
Decreasing the bin size down to $2 \ a$ does not change the trends noted below.
There is just a slow but steady decrease in $\langle d \rangle$ of only $\sim a$ that is nearly independent of configuration and time at $t_W/t_0 > 10^4$.

\begin{figure}
  \includegraphics[width=0.7\columnwidth]{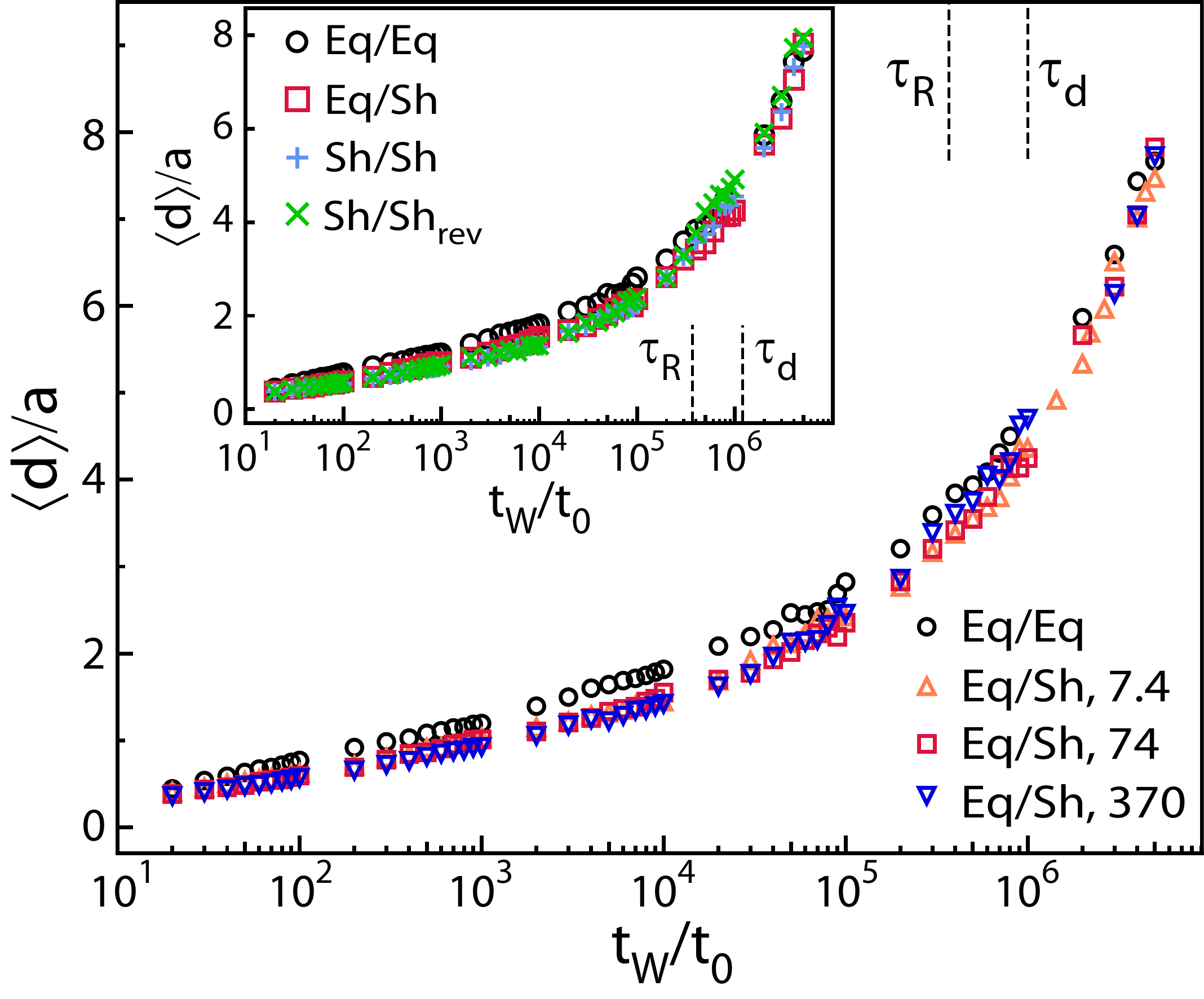}.
	\caption{Average interdiffusion distance $\langle d \rangle$ as a function of welding time for Eq/Eq and Eq/Sh systems sheared at the indicated $Wi_R$.
To reduce the effect of interface roughness, $\langle d \rangle$ is calculated relative to the mean interface position in squares of edge $10 \ a$.
Inset: Same quantity for different welding configurations with shear at $Wi_R=74$.
	}
  \label{fgr:avgd}
\end{figure}

Figure \ref{fgr:avgd} shows $\langle d \rangle = (\langle d_1\rangle + \langle d_2\rangle)/2$ as a function of welding time $t_W$ for the different systems analyzed.
All of the sheared systems show similar behavior, with little dependence on shear rate or configuration (inset).
The equilibrium systems show slightly greater penetration at times up to $\tau_R$, but the difference is less than $0.2 \ a$.
Note that this would imply faster diffusion of equilibrium systems while entanglement loss has been expected to produce faster diffusion of sheared systems.
If $\langle d \rangle$ is calculated without correcting for roughness,
equilibrium systems appear to interdiffuse by about the same amount as $Wi_R=7.4$ systems, and by up to $0.5 \ a$ less than other systems.
For all ways of measuring $\langle d \rangle$, variations with configuration and $Wi_R$ remain less than a molecular diameter ($\sim 0.5$ nm) and much less than the interface roughness.

While $\langle d\rangle$ is a measure of interdiffusion, past work shows that mechanical strength is most directly correlated with the formation of entanglements between chains that start on opposite sides of the interface.\cite{wool_polymer_1994,ge_structure_2013,ge_molecular_2013,ge_tensile_2014}.
Ge et al. \cite{ge_structure_2013, ge_molecular_2013, ge_tensile_2014}
applied PPA to welded states and determined the number of interfacial TCs, $N^I_{TC}$,
by counting contacts between the primitive paths of chains that started on different sides of the interface.
The shear strength and fracture energy of welds both rose linearly with the areal density of these interfacial entanglements until they saturated at the bulk strength.
This metric is independent of $z$ and thus not affected by interfacial roughness.

\begin{figure}[ht]
    \includegraphics[width=0.6\linewidth]{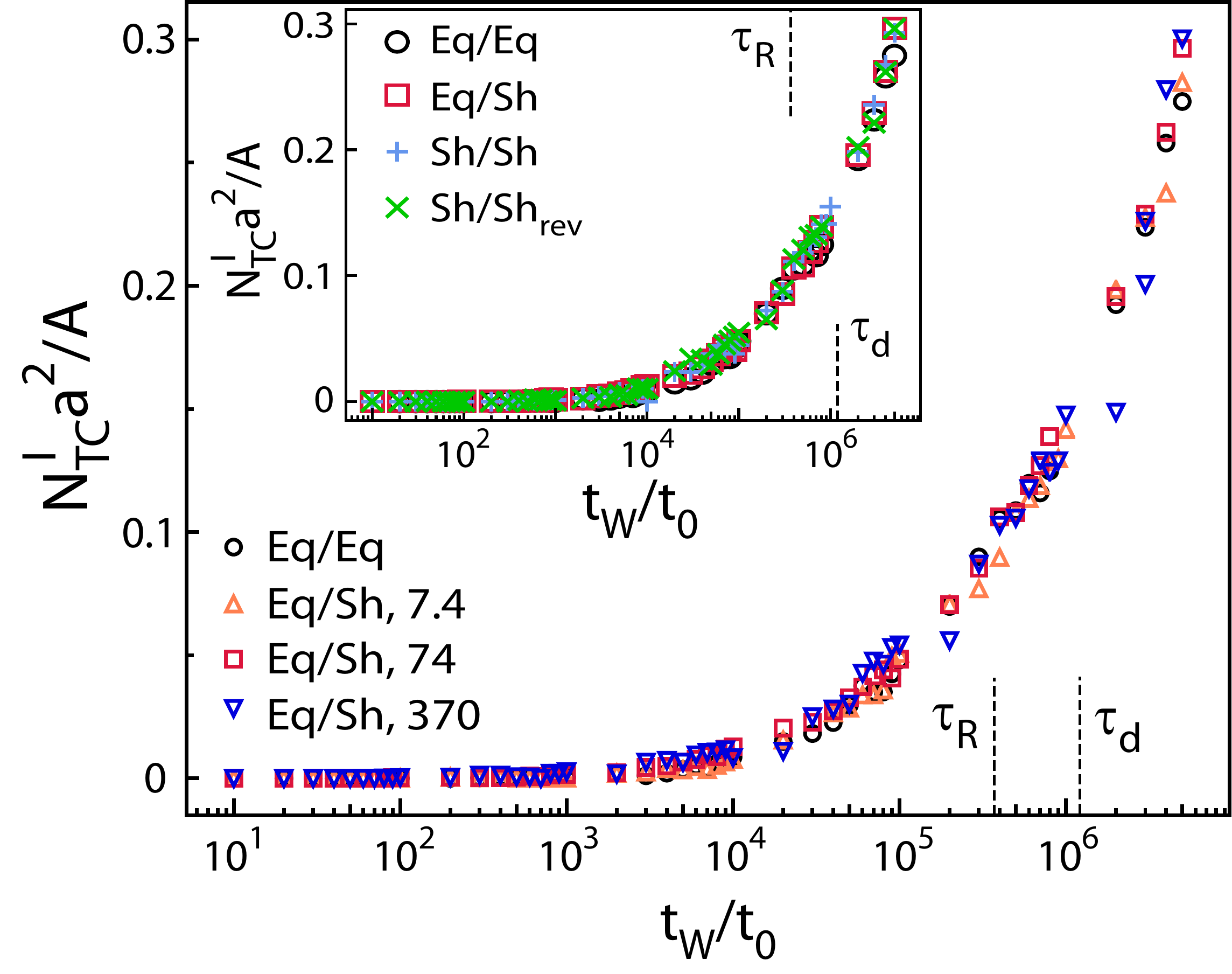}
    \caption{Areal density of interfacial entanglements as a function of $t_W$ for Eq/Eq and Eq/Sh systems sheared at the indicated $Wi_R$.
The area $A=L_x L_y$ and the Rouse and disentanglement times are indicated by vertical dotted lines.
Inset: Same quantities for Eq/Eq systems and different sheared configurations at $Wi_R=74$.
	}
    \label{fgr:NTC}
\end{figure}

Figure \ref{fgr:NTC} shows the areal density of interfacial entanglements from PPA as a function of time for different initial conditions.
The results for all starting states and shear rates are the same within statistical fluctuations.
This supports the conclusion that alignment does not affect the dynamics of mixing between molecules from different sides of the interface.
The broadening of the concentration profile for sheared systems at intermediate
times is just a roughening of the interface that
may have a minor effect on shear strength,\cite{rottler_molecular_2003}
but does not weld surfaces together.
Note that the most rapid change in Fig. \ref{fgr:ree} is contraction of chain ends along the $x-$direction, which does not drive diffusion.
The smaller expansion along $z$ is accommodated by fluctuations in the interface. It does not tend to push chain ends across the interface to form new entanglements.

Plotting the density of interfacial entanglements against $\langle d \rangle$ produces results very similar to Fig. 5 of Ge et al.\cite{ge_structure_2013}
There is little increase in $N_{TC}^I$ until $\langle d \rangle$ exceeds $\sim 1.5 \ a$.
This indicates that a minimum interpenetration is needed to form entanglements.
For larger $\langle d \rangle$, the areal density of entanglements rises linearly with $\langle d \rangle$.
The slope is consistent with that in Fig. 5 of Ge et al.\cite{ge_structure_2013} for all rates and configurations.

\subsection{Shear Strength }

The previous section showed that chain alignment had little effect on the rate of interdiffusion or entanglement formation.
We now show that initial states with different alignment still exhibit markedly different
mechanical properties.

\begin{figure}
  \includegraphics[width=0.45\columnwidth]{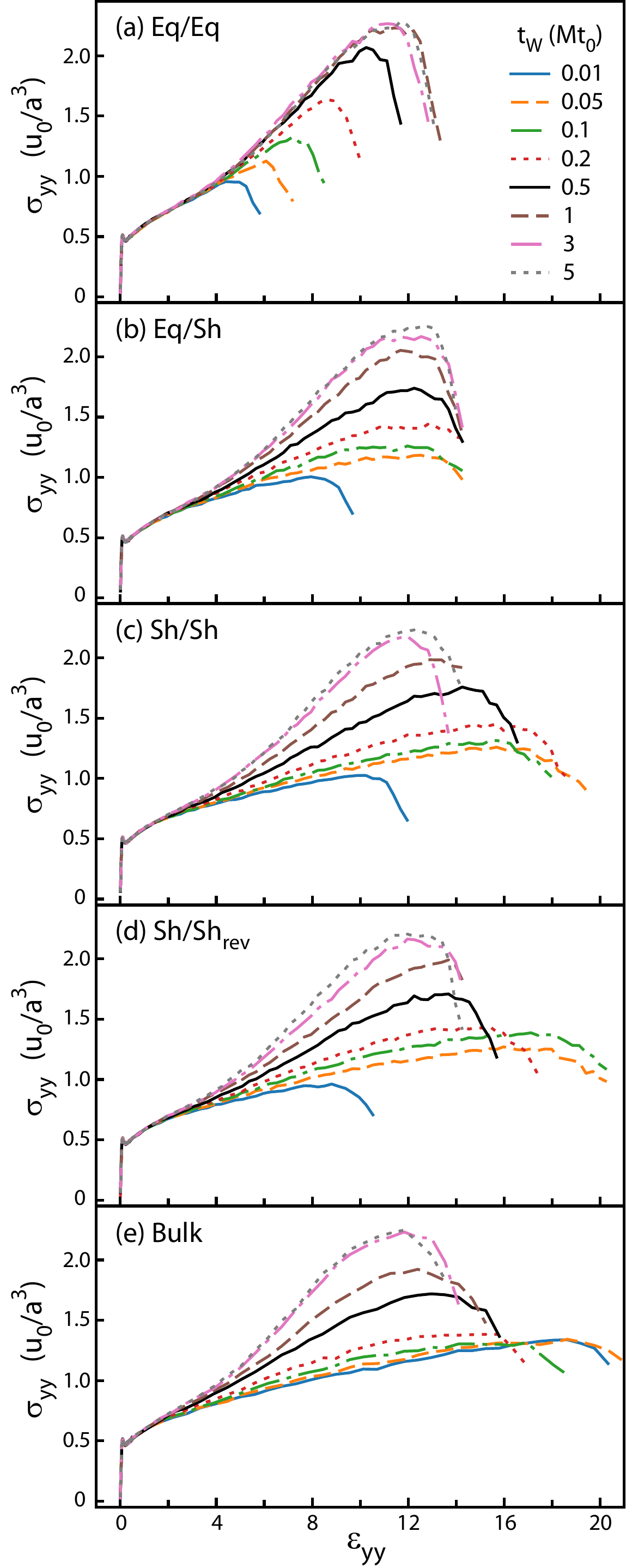}
	\caption{Shear stress $\sigma_{yy}$ as a function of the mean shear strain $\epsilon_{yy}$ for the different types of starting state:
	(a) Eq/Eq, (b) Eq/Sh, (c) Sh/Sh, (d) Sh/Sh$_{rev}$ and (e) bulk systems with no welded interface.
	Results are shown for the welding times in the legend of panel (a) and
	sheared slabs were aligned by steady-state shear with $Wi_R = 74$.
	}
  \label{fgr:shear}
\end{figure}

The shear strength of welds is normally characterized by the maximum shear stress the interface can withstand before failure, $\sigma_{max}$, which is measured experimentally with a lap joint shear test.\cite{wool_polymer_1994}
The variation of shear stress with shear strain $\gamma$ for simulations with different $t_W$ and initial configurations is shown in
Figure \ref{fgr:shear}.
As noted above, shear was applied along the interface and perpendicular to the printing direction to mimic tear tests of walls printed with FFF.\cite{seppala_weld_2017}

Results for unaligned Eq/Eq systems (Fig. \ref{fgr:shear}(a))
are consistent with past work.\cite{ge_structure_2013, ge_molecular_2013, ge_tensile_2014}
Curves for all weld times show the same initial yield stress ($\sim 0.5 \ a^3/u_0$) and strain hardening.
As $t_W$ increases, strain hardening extends to larger strains and higher stresses.
In this regime, the interface fails by chain pullout: The ends of chains that have diffused across the interface are pulled back to their starting side.
For sufficiently long weld times, $t_W  \geq 1 \ M t_0$, the entire curve becomes independent of time and consistent with bulk behavior.
In this regime, chains have diffused across the interface and entangled with other chains.
They break rather than being pulled back across the interface.
As in past work, bulk strength $\sigma_{max}^0$ is achieved at times where $N^I_{TC}/A \sim 0.15 \ a^{-2}$ and $\langle d \rangle \sim 4 \ a$.

\begin{figure}
  \includegraphics[width=0.6\columnwidth]{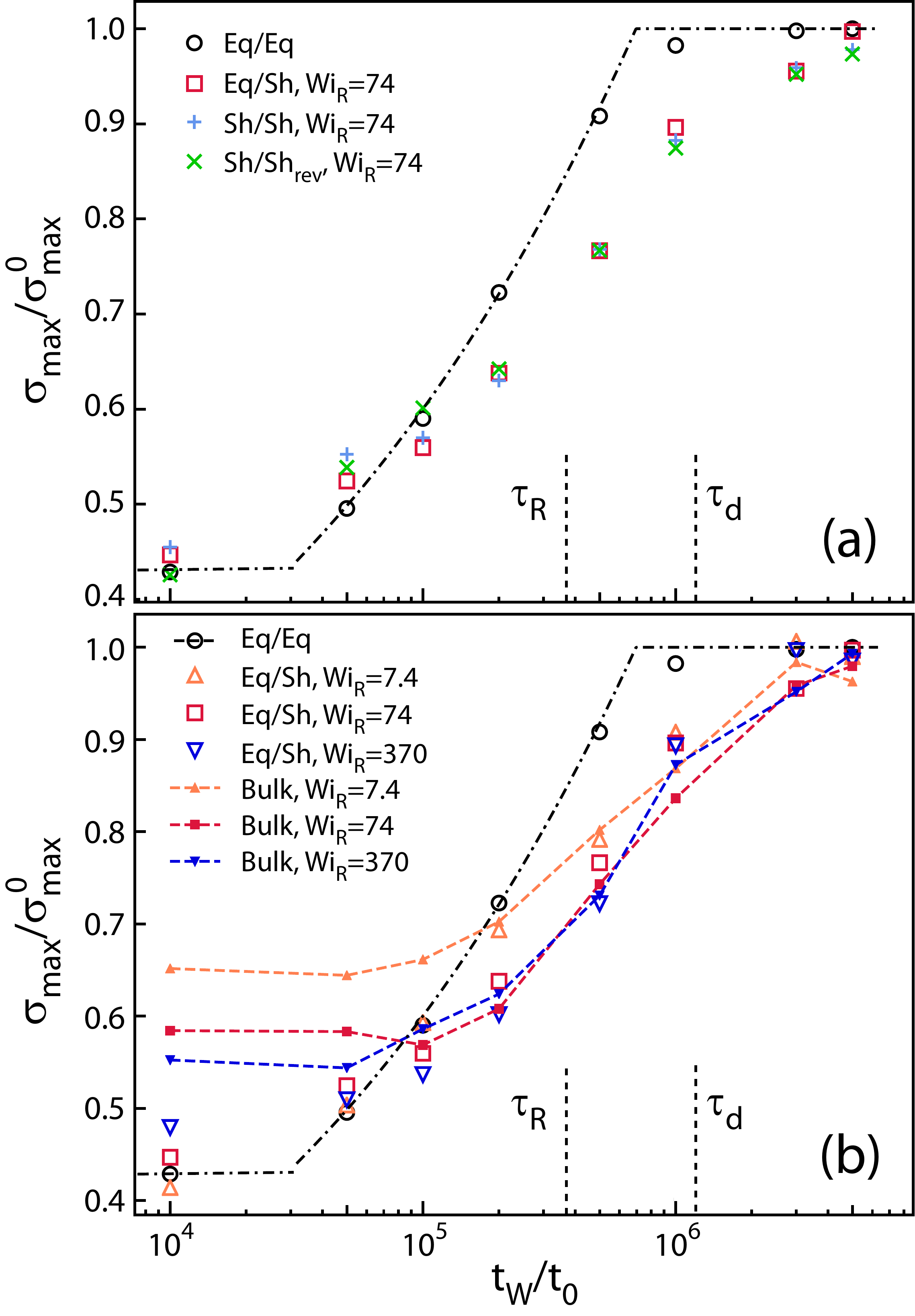}
	\caption{(a) Maximum shear stress normalized by the bulk equilibrium value $\sigma_{max}^0$ as a function of welding time for equilibrium systems (black) and different sheared states at $Wi_R=74$ (colors). A dot-dashed line shows a fit of Eq/Eq data to power law growth $\sim t^{1/4}$ followed by saturation at unity. (b) Rise in maximum stress with $t_W$ for Eq/Eq (black) and Eq/Sh systems at different $Wi_R$ (colors). Also shown are bulk results for the same initial shear alignments (small symbols with lines). Lines between points are a guide to the eye and vertical lines show the Rouse time and disentanglement time.
	}
  \label{fgr:smax}
\end{figure}

Aligned systems show somewhat different behavior. The most striking change is an increase in the strain to failure at small $t_W$. This reflects a more gradual strain hardening rather than an increase in maximum stress. During hardening, chains that are originally aligned along the $x$ direction are reoriented along $y$.
Because the chains are already stretched, this reorientation requires a relatively large strain.
The Eq/Sh system accommodates less strain than the other sheared systems because the unaligned half of the system does not deform as easily.
The ability of sheared systems to reorient is insensitive to the relative alignment of the two slabs (Sh/Sh or Sh/Sh$_{rev}$).

Alignment also changes the peak shear stress $\sigma_{max}$ and the time to reach it.
Note that $\sigma_{max}$ has saturated by $1 \ Mt_0$ for the Eq/Eq systems
in Fig. \ref{fgr:shear} but not until $5 \ M t_0$ for sheared systems.
Changes in the rate of strengthening are more clearly 
seen in Fig. \ref{fgr:smax}, which shows the time dependence of 
$\sigma_{max}$ normalized by its bulk unaligned value $\sigma_{max}^0$.
As in past simulations\cite{ge_molecular_2013} and experiments,\cite{wool_polymer_1994} Eq/Eq results can be fit to a power law rise $\sigma_{max}\sim t^{1/4}$
at intermediate times, followed by saturation at bulk strength.
This power law is motivated by models of reptation that predict $\langle d \rangle \sim t^{1/4}$ between $\tau_e$ and $\tau_d$ and assumes that the strength scales with $\langle d \rangle$.
In both experiment and simulations,\cite{ge_molecular_2013,wool_polymer_1994} there is a plateau in $\sigma_{max}$ at early times ($t_W< 5 \times 10^4 t_0$).
As noted above there is a minimum $\langle d \rangle$ required to form
entanglements\cite{ge_structure_2013} and the power law behavior only applies
once this threshold is passed.

Fig. \ref{fgr:smax}(a) shows that the strengths of all sheared configurations grow at a similar rate. They appear slightly stronger than Eq/Eq systems at very early times, perhaps because of the extra surface roughness.\cite{rottler_molecular_2003}
For $t_W > 10^5 t_0$ the strength of sheared systems grows more slowly and is just reaching $\sigma_{max}^0$ at the longest weld times.
Fig. \ref{fgr:smax}(b) shows samples aligned at all three $Wi_R$ exhibit similar behavior.
The depression of $\sigma_{max}$ increases with $Wi_R$ but the time to achieve bulk equilibrium strain is always about $5 Mt_0$.

\begin{figure}
  \includegraphics[width=0.45\columnwidth]{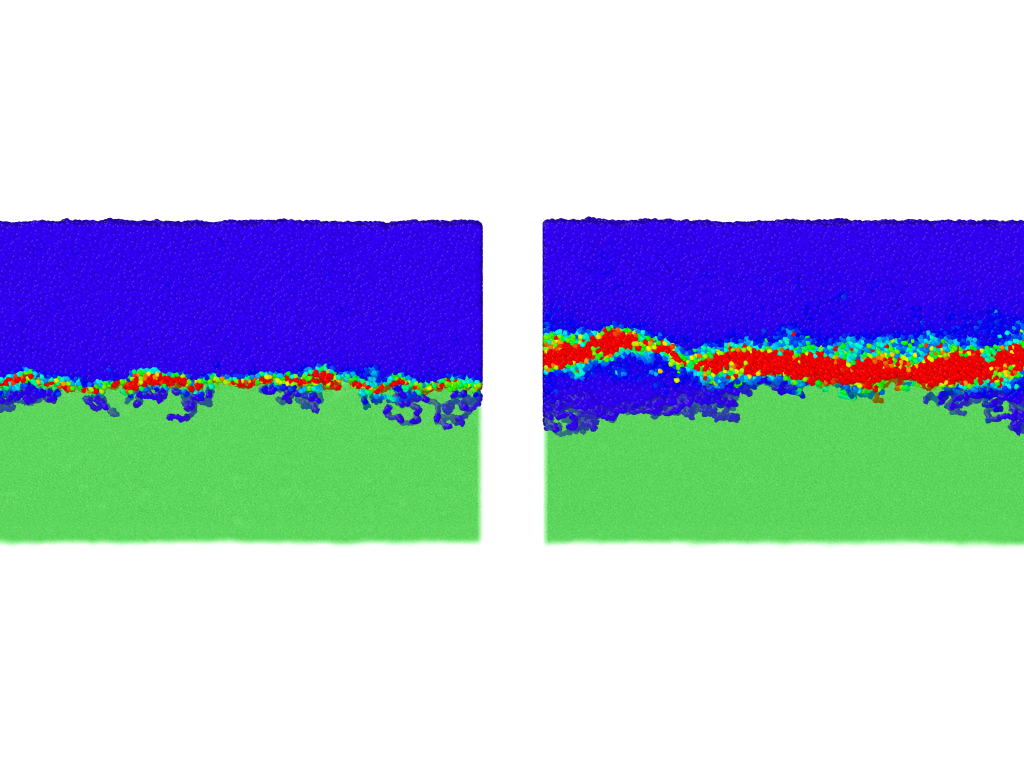}
  \caption{Pictures showing the strain is localized at the interface for Eq/Eq systems (left) and in the upper sheared material for Eq/Sh systems (right) at $t_W/t_0=4 \ Mt_0$.
Monomers from the upper surface are colored by the accumulated nonaffine strain with blue being low and red high. Monomers from the lower surface are a transparent green. One periodic image is shown in the $x-z$ plane.
Left: For Eq/Eq, red strained regions are at the height where $\phi(z) \approx 0$ even though some chains from the upper surface extend below this interface. For this $t_W$ they would be pulled back to the other side at larger strains.
Right: For the Eq/Sh system the interface between Eq and Sh is noticeably rougher and sheared atoms are all in the upper half, particularly at left of the image.
	}
  \label{fgr:deform}
\end{figure}

The slow rise in $\sigma_{max}$ for sheared systems is surprising given that they show the same rate of growth in $N_{TC}^I$.
The explanation is that the strength of sheared systems is not limited by the interfacial strength of welds, but by the alignment of adjacent bulk regions.
To demonstrate this we show results for bulk systems without an interface.
The stress strain curves in Fig. \ref{fgr:shear}(e) are very similar to those
for the sheared systems in panels (c) and (d).
The only statistically significant difference is that the bulk system has a larger yield strain and stress at the earliest times. Dashed lines in Fig. \ref{fgr:smax}(b) show that the time dependent growth in $\sigma_{max}$ for bulk aligned systems is similar to that for welded systems aligned at the same $Wi_R$. The only statistically significant difference is at $t_W \leq 3 \times 10^4 t_0$ where the bulk system is stronger.
As illustrated in Fig. \ref{fgr:deform},
direct examination of welded systems confirms that plastic deformation is localized in sheared regions and not at the welded interface.

\subsection{Fracture Energy}

\begin{figure}
  \includegraphics[width=0.45\columnwidth]{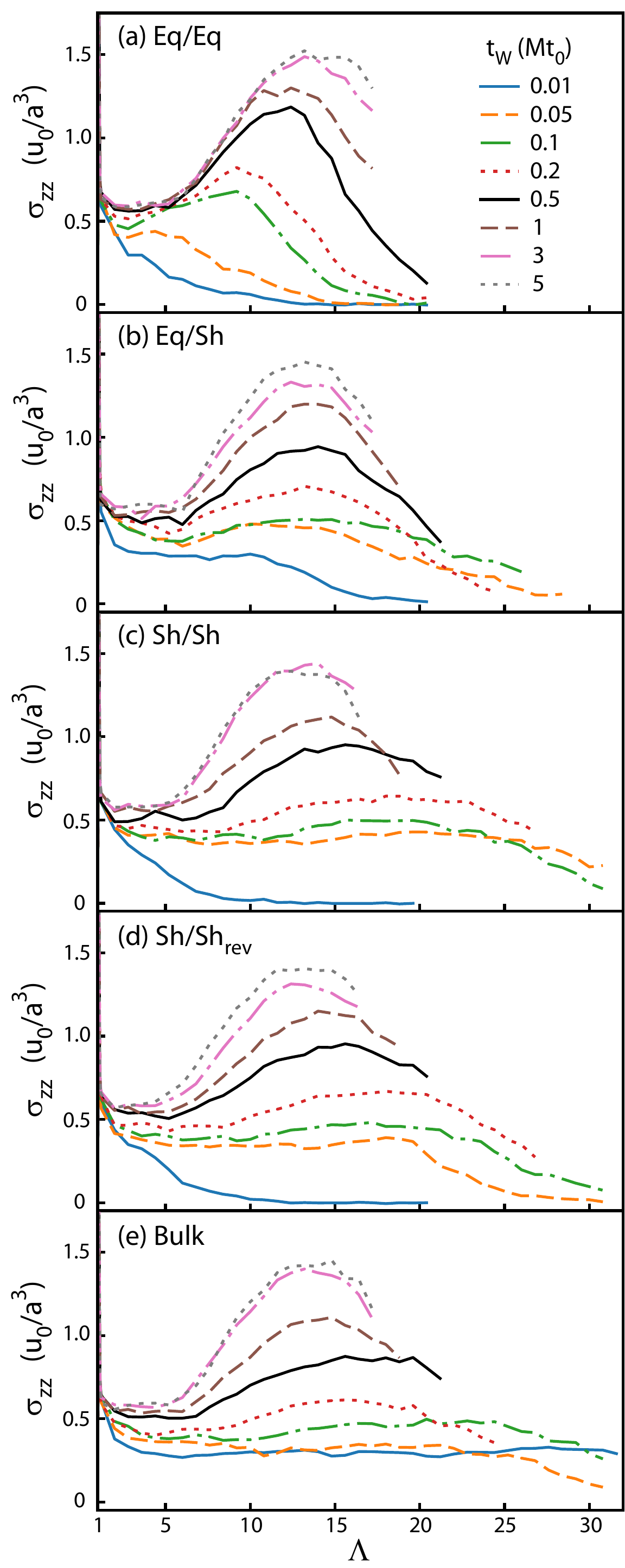}
	\caption{Tensile stress $\sigma_{zz}$ as a function of the stretch $\Lambda$ along $z$ for the different types of starting state:
	(a) Eq/Eq, (b) Eq/Sh, (c) Sh/Sh, (d) Sh/Sh$_{rev}$ and (e) bulk systems with no welded interface.
	Results are shown for the welding times in the legend of panel (a).
	Sheared slabs were aligned at $Wi_R = 74$ and $\Lambda$ is the ratio of the separation between rigid walls to its initial value.
	}
  \label{fgr:tensile}
\end{figure}

Figure \ref{fgr:tensile} shows the variation of tensile stress $\sigma_{zz}$ with the expansion of the system $\Lambda$ defined as the ratio of the spacing between rigid walls to the initial undeformed value.
Once again, the Eq/Eq results are consistent with past work.\cite{ge_tensile_2014}
All systems show an initial peak of around $2 \ u_0/a^3$ where cavities begin to form. This peak occurs at $\Lambda$ very near unity and lies essentially on the vertical axis of the figure.

For the shortest $t_W$, $\sigma_{zz}$ drops monotonically from this peak value.
In such cases,
the fracture energy per unit area $G_i$ can be obtained by integrating the stress times the displacement of walls.
For longer times a plateau becomes visible for $\Lambda$ up to about 7.
This plateau is the signature of craze formation, where 
the initial dense polymer is converted into a network of oriented fibrils and voids.\cite{haward_1997,ward_2012}
For $0.05Mt_0$ the interface fails through chain pullout after only part of the 
volume has been converted into a craze.
At larger $t_W$ the entire volume is converted to a craze by about $\Lambda=7$.
The stress needed to deform the craze grows with $\Lambda$ until it fails.
Chain pullout is observed for small times but for $t_W \geq 3 Mt_0$ chains break and the strength approaches the bulk value.

Figures \ref{fgr:tensile} (b-e) shows how alignment affects craze formation.
Once again, alignment increases the deformation required to produce failure,
and the increase is larger when both slabs are aligned and contribute to the deformation.
In addition,
results for Sh/Sh and Sh/Sh$_{rev}$ systems are not significantly different.
Figure \ref{fgr:tensile}(e) shows that these changes predominantly reflect the response of oriented bulk regions rather than the interface.
The main difference between bulk and welded systems is at the very earliest time $\sim 10^4 t_0 \sim t_e$ where the welded systems do not show a plateau while the bulk system has the longest plateau.
At all longer times the location of failure in welded systems moves from the interface to oriented bulk regions.

Following past work,\cite{rottler_cracks_2002,ge_tensile_2014}
Brown's model\cite{brown_molecular_1991} can be used to determine the fracture
energy of systems that craze.
Brown approximates the fracture as crack propagation through an elastic
crazed region between rigid uncrazed regions.
One obtains:
\begin{equation}
	G_i = 4\pi\kappa S (S_{max}/S)^2 D_0 (1-1/\Lambda_{max})
    \label{eq:brown}
\end{equation}
where $\kappa$ reflects anisotropy in the elastic properties of the craze,
$S$ is the plateau stress seen in Fig. \ref{fgr:tensile}, $S_{max}$ is the peak tensile stress, $D_0$ is the distance between craze fibrils and $\Lambda_{max}$ is the stretch of regions as they are deformed into a craze.
As in Ge et al.,\cite{ge_tensile_2014}
we use typical values of $D_0=12 a$, $\kappa=2.5$ and $\Lambda_{max}=8$ to calculate $G_i$.
The observed value of $\Lambda_{max}$
increases slightly with alignment, but this has little effect on the toughness ($<10$\%).
Estimates of changes in $D_0 \sim 12a$ and $\kappa \sim 2.5$ with the alignment of the initial state are also small.

\begin{figure}
  \includegraphics[width=0.6\columnwidth]{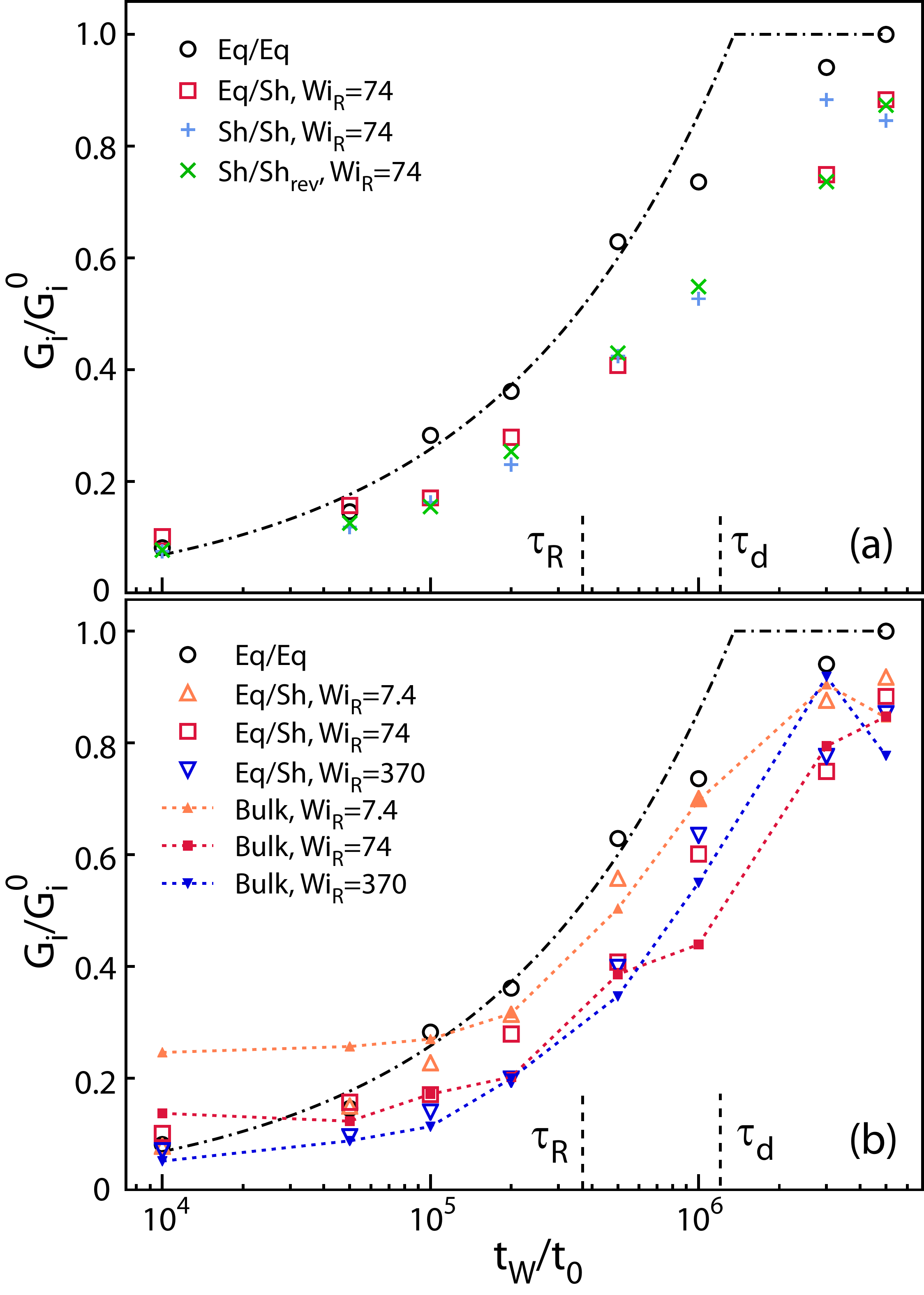}
	\caption{(a) Tensile fracture energy normalized by bulk equilibrium value $G_i^0$ as a function of welding time for different configurations.
	(b) Rise in tensile fracture energy for different initial alignments.
	Large symobols are for welded systems and smaller symbols connected by broken lines are for bulk systems with the same alignment.
The dash-dotted line in both panels is a fit to $G_i \propto t_W^2$ for $G_i < G_i^0$.
	Vertical dashed lines indicate $\tau_R$ and $\tau_d$.
	}
  \label{fgr:Gi}
\end{figure}

Figure \ref{fgr:Gi}(a) shows the tensile fracture energy for systems at equilibrium or sheared at $Wi_R=74$. For all systems, $G_i$ rises by an order of magnitude before saturating at the bulk value.
In experiments, the initial rise is often fit to a power law $G_i \propto t_W ^{1/2}$.
A similar fit to Eq/Eq results is shown by a dashed line.
All sheared systems show nearly the same behavior
and rise more slowly than equilibrium systems.
Even at the longest times, they have not reached bulk strength.
Figure \ref{fgr:Gi}(b) shows that strength reduction rises as $Wi_R$ increases.

As for $\sigma_{max}$, the reduction in $G_i$ with increasing alignment
is due to a decrease in bulk strength rather than weaker interfacial welds.
Small symbols connected by lined
in Fig. \ref{fgr:Gi}(b) show the fracture energy of bulk systems without any interface. 
For each $Wi_R$ these curves lie slightly above welded systems at short times
($<10^5 t_0$) but are statistically equivalent to results for welded systems at long times.
Direct evaluation of molecular conformations confirms that the craze is confined to the sheared portion of deforming welded systems.

\section{Discussion and Conclusions}

The results presented above have important implications for optimizing FFF protocols. First, they imply that the rate of diffusion to form welds and the time for molecules to relax from stretched aligned states are described by equilibrium relaxation times from tube theory. The same result is found in our ongoing bulk simulations of chains with different lengths and entanglement lengths, and similar results have been found for elongational flow.\cite{oconnor_stress_2019}
Past experimental studies have extracted an effective welding time from the thermal history of weld lines and time-temperature superposition.\cite{seppala_weld_2017,mcilroy_deformation_2017,mcilroy_disentanglement_2017,coogan_prediction_2020}
Figures \ref{fgr:ree} and \ref{fgr:P2t} imply that this analysis need not
include any dependence of relaxation times on alignment.
They also provide further evidence that improved models are needed for describing entanglements and their evolution in highly aligned states.
\cite{ianniruberto_convective_2014,baig_flow_2010,milner_microscopic_2001,nafar_sefiddashti_individual_2019,mcilroy_deformation_2017,mcilroy_disentanglement_2017,oconnor_stress_2019}

The fact that the rate of interfacial entanglement formation is independent of alignment (Fig. \ref{fgr:NTC}), implies that the rate at which welds strengthen should also be described by equilibrium dynamics.
Welds should then be as strong as isotropic bulk material after molecules have diffused along their tube by only a couple of entanglement lengths.\cite{ge_structure_2013, ge_molecular_2013, ge_tensile_2014}
However, fabricated parts are only as strong as the weakest link
and alignment reduces the resistance of bulk material to shear and fracture.
The strength of initially aligned material near the weld grows until
several times $\tau_d$, the time needed to diffuse out of the entire length of the tube.
As predicted by tube theory, molecular orientation relaxes exponentially with a time $\tau_d$
and strength only saturates when almost all orientation is lost.
This criterion can be incorporated in future theories for the evolution of order
and strength in FFF. The immediate implication is that to achieve bulk strength printing parameters should be adjusted to prevent significant alignment or the welding time must exceed several $\tau_d$ so that any initial alignment is lost.

The effect of alignment on mechanical strength has been studied in bulk glassy polymers with both experiment and simulation.\cite{ge_anisotropic_2010,haward_1997,senden_strain_2010,curtis_effect_1971,miller_cleavage_1971}
The resistance to extension along the chain backbone is increased, while the transverse directions become weaker.
Unfortunately, these weaker directions are involved in tensile fracture at FFF welds and shear transverse to the print direction.
Parts may be stronger along the print direction, but this 
is not as important to performance.
Deposition produces the greatest alignment at the weld interface, and parts are generally thinnest at the weld lines.
These factors combine to localize failure near welds even if the few nanometers where molecules have welded by interdiffusion is not weaker than the bulk.

\begin{figure}
  \includegraphics[width=0.6\columnwidth]{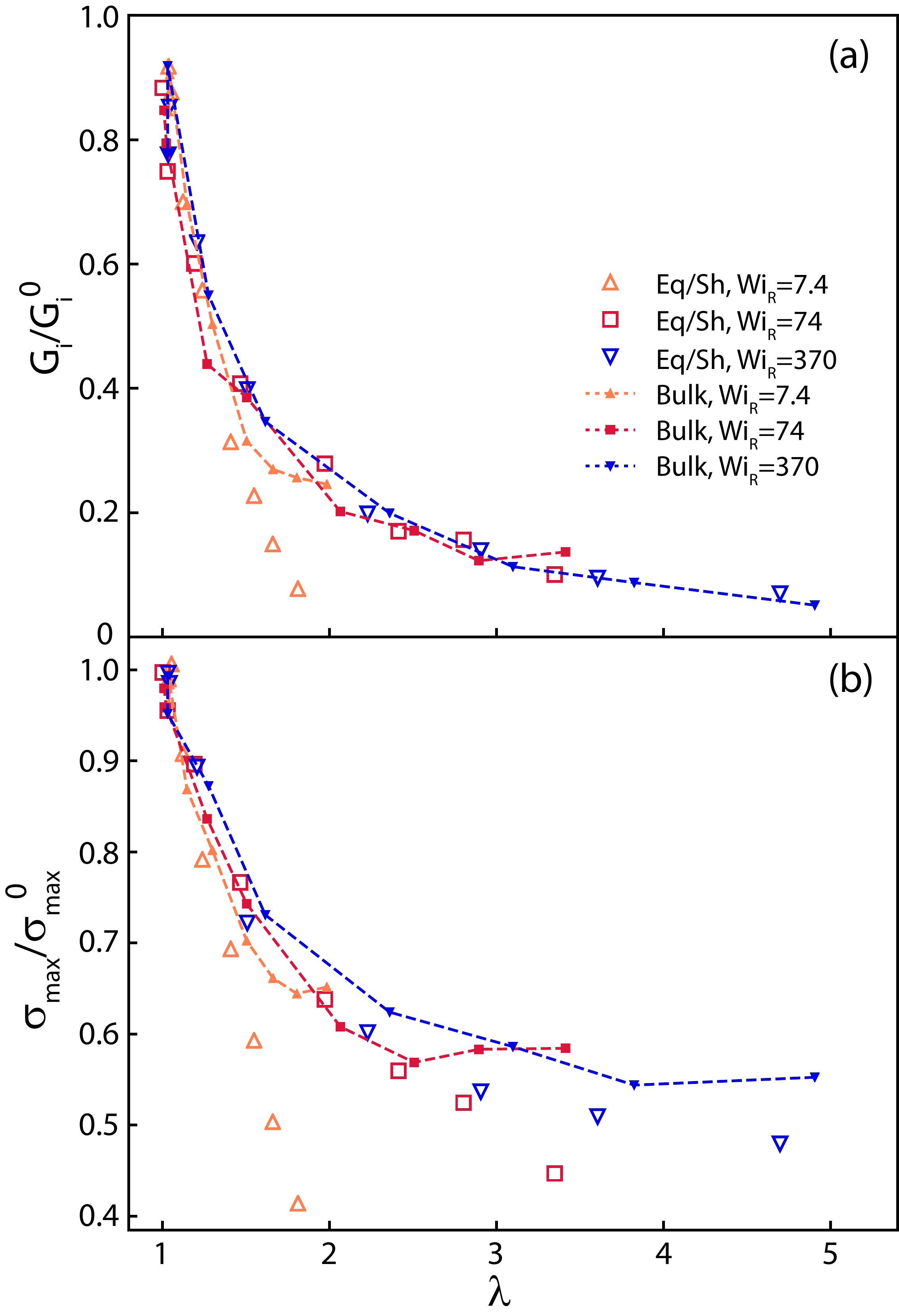}
	\caption{(a) Tensile fracture energy and (b) maximum shear stress as a function of the average end-end stretch of chains $\lambda$.
	As the stretch in welded systems relaxes, the strength approaches the bulk curves for aligned systems (dashed lines).
}
  \label{fgr:Givslam}
\end{figure}

At macroscopic scales, orientation is often measured with birefringence, an anisotropy in the index of refraction that correlates with the orientation and stretch $\lambda$ of individual molecules.
Plots of the tensile fracture energy of glassy polymers such as PMMA, PS and PC 
against birefringence show the fracture energy for crack propagation along the alignment axis drops with increasing orientation and then saturates.\cite{curtis_effect_1971,miller_cleavage_1971}
Figure \ref{fgr:Givslam} shows $\sigma_{max}$ and $G_I$ plotted against the
end-end stretch $\lambda$.
Bulk systems (dashed lines) follow a common curve that
is qualitatively similar to experiments.
Strength drops rapidly with increasing orientation and then saturates.
Welded systems can not be stronger than the corresponding bulk and thus lie below the dashed lines. 
At very early times the weld is weak and failure occurs there.
At longer times the weld is irrelevant and $\sigma_{max}$ and $G_i$ follow the bulk curves.

The coarse-grained model used here has successfully described general features of the mechanical response of glassy polymers and their connection to entanglements.
It is expected to capture the behavior of commonly used materials such as polycarbonate (PC) and poly-lactic acid (PLA) that remain amorphous under most printing conditions.
Chemically detailed models are needed to make quantitative predictions of strength,
but are difficult to use on the time scales needed for highly entangled polymers
to interdiffuse.\cite{luchinsky_welding_2020}
Different mechanisms may be important for
semicrystalline polymers or polymers with multiple monomers, such as
acrylonitrile butadiene styrene (ABS).
Rather than being constrained by entanglements, the relative motion of chains may be constrained by crystal order or segregation of monomer species.
Both effects will be interesting topics for future study.

\begin{acknowledgement}

This material is based upon work supported by the National
	Science Foundation under Grant No. NSF DMREF-90069795 and by Fellowship 235249/2014-9 from the National Council for Scientific and Tecnological Development (CNPq) of Brazil. We thank Martin Kroger for supplying the Z1 code, and C. McIlroy, K. Migler, T. D. Nguyen, T. C. O'Connor, P. Olmsted, and J. Seppala for useful discussions. Simulations were performed at the Maryland Advanced Research Computing Center.

\end{acknowledgement}

\bibliography{welding}

\end{document}